\documentclass[aps,prl,floats,twocolumn,floatfix,showpacs]{revtex4}
\usepackage{epsf,graphicx}
\usepackage{amssymb}
\usepackage{amsmath}
\usepackage{eepic}
\usepackage[dvips]{color}
\newlength{\piclen}
\begin{document}

\title{Magnetic excitations in Kondo liquid: Superconductivity and Hidden Magnetic Quantum Critical Fluctuations}

\author{Yi-feng Yang$^{1}$, Ricardo Urbano$^{1,2}$, Nicholas J. Curro$^{3}$ and David Pines$^{3}$}
\affiliation{$^1$Los Alamos National Laboratory, Los Alamos, New Mexico 87545}
\affiliation{$^2$ National High Magnetic Field Laboratory, Florida State University, Tallahassee, Florida 32306-4005}
\affiliation{$^3$Department of Physics, University of California, Davis, California 95616}
 \date{September 7, 2009}

\begin{abstract}
We report Knight shift experiments on the superconducting heavy electron material CeCoIn$_5$ that allow one to track with some precision the behavior of the heavy electron Kondo liquid in the superconducting state with results in agreement with BCS theory. An analysis of the $^{115}$In nuclear quadrupole resonance (NQR) spin-lattice relaxation rate $T_1^{-1}$ measurements under pressure reveals the presence of 2d magnetic quantum critical fluctuations in the heavy electron component that are a promising candidate for the pairing mechanism in this material. Our results are consistent with an antiferromagnetic quantum critical point (QCP) located at slightly negative pressure in CeCoIn$_5$ and provide additional evidence for significant similarities between the heavy electron materials and the high $T_c$ cuprates. 
\end{abstract}


\maketitle

A central issue in heavy electron physics is the origin of unconventional superconductivity (SC) at low temperatures, and the role played by quantum critical fluctuations in bringing that about. Valuable information on both is provided by nuclear magnetic resonance (NMR) experiments of the Knight shift and spin-lattice relaxation rates \cite{Curro2009}, but as is so often the case the devil is in the details - in this case being able to separate out the contributions made by local moments to both quantities, so that one can be confident one is measuring the behavior of the heavy electrons alone and not a combination of the two. We are now in a position to do this, thanks to a phenomenological two-fluid description that treats the emergence of heavy electrons as the formation of a distinct new quantum state of matter that displays universal behavior below a characteristic temperature $T^*$ \cite{Nakatsuji2004,Curro2004,Yang2008a,Yang2008b,Yang2009}. The growth of this new phase is accompanied by a loss in strength of the spectral weight of the local moment contribution, and both phenomena are characterized by an order parameter, $f(T)$, so that it becomes possible, for a given physical phenomenon, to track the role played by residual local moments and focus on the behavior of the emergent phase, the heavy electron Kondo liquid.

In this communication, we first apply the two-fluid description to new measurements of the Knight shift in CeCoIn$_5$, and show that it provides a natural explanation for the anomalous behavior of the NMR Knight shift that is seen in both the normal and superconducting phases. The Knight shift anomaly, defined as the deviation of the total Knight shift from the magnetic susceptibility, shows scaling behavior in both phases, is independent of the probe nuclei, follows the predicted Kondo liquid universal behavior in the normal phase \cite{Nakatsuji2004,Curro2004,Yang2008a}, and BCS theory in the superconducting phase \cite{Yosida1958}, thereby establishing unambiguously that the heavy electron SC originates in the condensation of the heavy Kondo liquid.

We then carry out a two-fluid analysis of the NQR measurements of the $^{115}$In spin-lattice relaxation rate in the normal state of CeCoIn$_5$ \cite{Kohori2006}, and find that for pressures up to 1.2$\,$GPa, the heavy electron spin-lattice relaxation rate, $T_{1h}^{-1}$, takes the simple form expected for 2d magnetic quantum critical fluctuations, $T_{1h}T\propto T+T_0$ \cite{Barzykin2009}. This behavior persists up to $T^*$, while the offset, $T_0$, tends toward zero as one moves toward a candidate QCP at a slightly negative pressure \cite{Sidorov2002}. Such behavior is hidden in the overall spin-lattice relaxation rate by the presence of a substantial contribution from the unhybridized local moments, and can only be identified once this component is subtracted. 

\begin{figure}[t]
{\includegraphics[width=9 cm,angle=0]{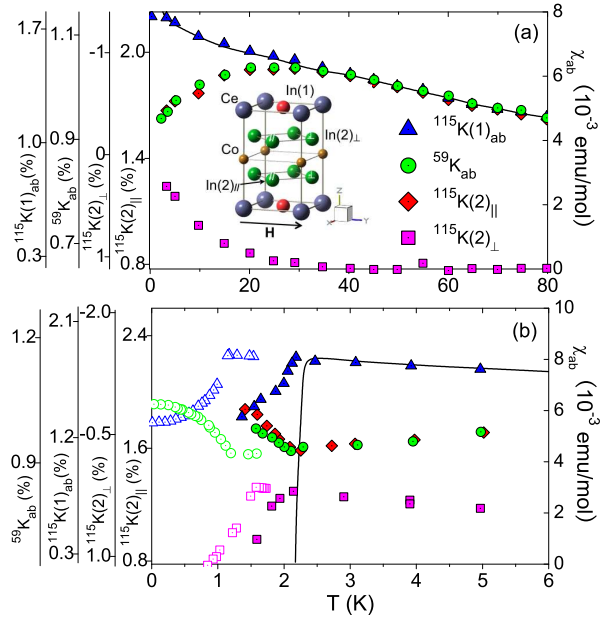}}
\caption{(Color online)
{Planar Knight shift of CeCoIn$_5$ at different probe nuclei in both normal (a) and superconducting (b) phases. The solid lines are the measured magnetic susceptibility. The solid symbols denote previous experimental data from Ref.~\cite{Curro2001} and the open symbols are our new data. The difference in $T_c$ is due to the different external magnetic field for the measurements. The inset indicates the position of the different probe nuclei (In(1), Co, In(2)$_\perp$ and In(2)$_\parallel$) in the unit cell. There are two inequivalent In(2) sites, depending on whether the field is parallel (In(2)$_\parallel$) or perpendicular (In(2)$_\perp$) to the unit cell face.}
\label{Fig:KS}}
\end{figure}

\begin{figure}[t]
{\includegraphics[width=9cm,angle=0]{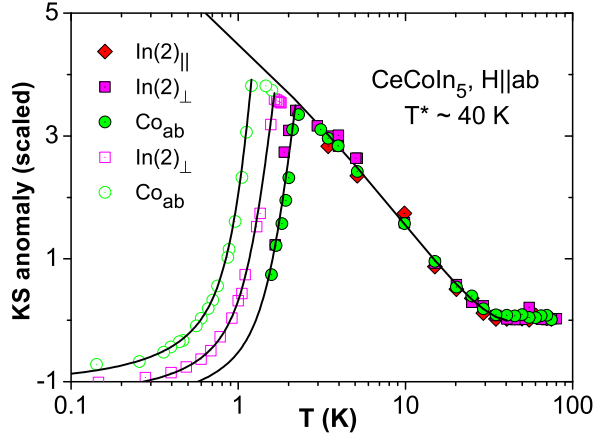}}
\caption{(Color online)
{Planar Knight shift anomaly of CeCoIn$_5$ at different probe nuclei in both normal and superconducting phases. The solid lines are fit to the universal Kondo liquid density of states in the normal phase \cite{Yang2008a} and BCS theory in the superconducting phase \cite{Yosida1958}. The best fit gives a superconducting gap $\Delta(0)/T_c\sim 4.5$ and a specific heat jump $\Delta C/C\sim 4$, both in good agreement with previous estimates \cite{Kohori2006,Sidorov2002}.}
\label{Fig:KSanomaly}}
\end{figure}

We have performed NMR measurements on CeCoIn$_5$ down to 0.04 K and the results in Fig.~\ref{Fig:KS} complement our previous measurements \cite{Curro2001}. According to the two-fluid scenario \cite{Curro2004,Yang2008a}, the magnetic susceptibility and Knight shift are related through
\begin{eqnarray}
\chi &=& \chi_h + \chi_l,\nonumber\\
K - K_0 &=& A \chi_h + B \chi_l,
\end{eqnarray}
where $\chi_h$ is the heavy electron susceptibility of the Kondo liquid and $\chi_l$ is the contribution to the susceptibility from the residual unhybridized local moments. $A$ and $B$ are the hyperfine couplings of the probe nucleus to the two components, respectively. $K_0$ is a constant offset that accounts for other contributions such as the Pauli term from the "light" conduction electrons. Above $T^*\sim 40\,$K, $\chi_h = 0$ and the Knight shifts measured for the In(1), Co, and In(2) sites are proportional to the magnetic susceptibility, $K = K_0 + B \chi$. Below $T^*$, the emergence of $\chi_h$ leads to the deviation from this linear relation shown in Fig.~\ref{Fig:KS}(a). The Knight shift anomaly is defined as
\begin{equation}
K_{anom} = K - K_0 - B \chi = (A-B)\chi_h,
\label{Eq:anom}
\end{equation}
and provides a direct measurement of $\chi_h$. An analysis of the Knight shift anomaly in many heavy electron materials has led to the conclusion that $\chi_h$ displays universal behavior \cite{Curro2004,Yang2008a}.  In Fig.~\ref{Fig:KSanomaly}, the planar Knight shift anomalies at these probe nuclei are found to be in good agreement with the predicted universal Kondo liquid density of states \cite{Yang2008a}, $\rho_h(T)=(1-T/T^*)^{3/2}[1+\ln(T^*/T)]$.

We consider next the extension of the two-fluid scenario to the superconducting phase. The planar Knight shift of CeCoIn$_5$ has several unexpected features that are not present in most other heavy electron materials and find a simple explanation within the two-fluid scenario. First, no anomaly is observed in the planar Knight shift data at the In(1) site. According to Eq.~(\ref{Eq:anom}), this suggests a cancellation of the In(1) hyperfine couplings to the residual local moments and the heavy Kondo liquid. Since the superconducting transition does not change the hyperfine couplings, this cancellation is expected to hold across $T_c$. Therefore, the In(1) nucleus probes the total spin susceptibility in the whole temperature range. This allows us to determine the intrinsic spin susceptibility, $\chi=(K-K_0)_{\text{In}(1)}/B_{\text{In}(1)}$, even below $T_c$ where the measured susceptibility goes negative. So a Knight shift anomaly can still be defined in the superconducting phase for probe nuclei other than In(1) by using Eq.~(\ref{Eq:anom}). In general cases where $A\neq B$ for all probe nuclei, the Knight shift anomaly or $\chi_h$ can be obtained (up to a constant prefactor) in a similar way even though the effective susceptibility defined as $(K-K_0)/B$ does not give the true spin susceptibility $\chi$.

Second, the planar Knight shift at the In(2)$_\perp$ site is a constant above $T^\star$, but evolves with temperature below $T^*$. This indicates that the In(2)$_\perp$ nucleus is not coupled to the localized moments. This special feature allows us to identify unambiguously the Kondo liquid behavior in both the normal and superconducting phases and provides an independent check on the Knight shift anomaly measured by other probe nuclei (In(2)$_{\parallel}$ or Co). As can be seen in Fig.~\ref{Fig:KSanomaly}, the In(2)$_\perp$ Knight shift anomaly above $T_c$ does fall upon the universal curve for the Kondo liquid after performing a simple scaling. Without any artificial manipulation, the Knight shift data at In(2)$_\perp$ nucleus provides added confirmation for the existence of the predicted universal Kondo liquid state.

Fig.~\ref{Fig:KS}(b) compares our new experimental data in the superconducting phase with that obtained earlier \cite{Curro2001}. By using the effective susceptibility obtained from the Knight shift at In(1) nucleus, the scaled Knight shift at In(2)$_\perp$, and the subtracted Knight shift anomalies at In(2)$_\parallel$ and Co are compared in Fig.~\ref{Fig:KSanomaly} for the whole temperature range below $T^\star$. Due to the difference in the applied external magnetic field, the superconducting transition temperatures are slightly different. However, the similar scaling behavior at different probe nuclei supports the Kondo liquid scenario in both the normal and superconducting phases and shows distinctly the condensation of the Kondo liquid into the superconducting state. Indeed, the rapid decrease of the anomalies follow exactly the BCS prediction for d-wave superconductivity \cite{Yosida1958}, 
\begin{equation}
K_{anom}(T)-K_{anom}(0)  \propto \int\,dE\left(-\frac{\partial f(E)}{\partial E}\right)N(E),
\end{equation}
where $f(E)$ is the Fermi distribution function and $N(E)\propto \langle |E|/\sqrt{E^2-\Delta_k(T)^2}\rangle_{FS}$ is the BCS density of states. The superconducting gap function is described by the usual interpolation formula 
\begin{equation}
\Delta_k(T)=g_k\Delta(0)\tanh\left[\sqrt{\left|\frac{\partial \Delta^2}{\partial T}\right|_{T_c}\frac{T_c}{\Delta(0)^2}\left(\frac{T_c}{T}-1\right)}\,\right],
\label{Eq:BCS}
\end{equation}
where $\Delta(0)$ is the gap amplitude and $g_k$ describes the gap symmetry. In Fig.~\ref{Fig:KSanomaly}, the best fit for d-wave gives $\Delta(0)/T_c\sim 4.5$  and a specific heat jump $\Delta C/C\sim 4$, both in good agreement with previous estimates \cite{Kohori2006,Sidorov2002}. We therefore conclude that the Kondo liquid is responsible for the heavy electron superconductivity. 

The extension to the superconducting state reveals a third unexpected feature, a negative Knight shift anomaly at zero temperature, which is best understood in the In(2)$_\perp$ data. Above $T^*$, In(2)$_\perp$ probes only conduction electrons that have yet to hybridize. These electrons contribute a constant term in $K_0$ in the normal phase. Therefore, we surmise that the negative anomaly relative to its high temperature offset $K_0$ originates in the loss of these conduction electrons, in addition to the heavy electron $\chi_h$, to the superconducting condensate. A similar loss of their contribution is expected for the other probes, In(2)$_\parallel$ and Co, and its presence in Fig.~\ref{Fig:KSanomaly} confirms this physical picture. Since the In(2)$_\parallel$ and Co Knight shift anomalies are obtained after subtracting an effective susceptibility deduced from the In(1) data, this picture also explains the upturn of the In(2)$_\parallel$ and Co Knight shifts below $T_c$ and their consequential intersection with the In(1) Knight shift seen in Fig.~\ref{Fig:KS}(b).

\begin{figure}[t]
{\includegraphics[width=9cm,angle=0]{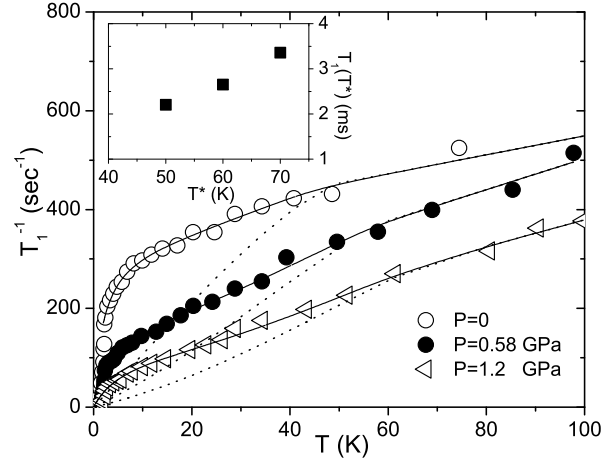}}
\caption{
{$^{115}$In NQR relaxation rate of CeCoIn$_5$ at different pressures analyzed by the two-fluid scenario. The data are reproduced from \cite{Kohori2006}. The solid lines are the best fit to the total $T_1^{-1}$ and the dotted lines show the local moment contribution. The inset compares the magnitude of $T_1$ at $T^*$ with the exchange interaction $T^*$.
}
\label{Fig:CoT1}}
\end{figure}

Important information on the spin fluctuation excitations that may be responsible for the measured d-wave pairing may be obtained from the measured $^{115}$In NQR spin-lattice relaxation rate $T_1^{-1}$ of CeCoIn$_5$ under pressure \cite{Kohori2006} shown in Fig.~\ref{Fig:CoT1}. Since, however, below $T^*$ both local moment and heavy electron spin fluctuations contribute to the overall relaxation rate, it is essential to separate these out before drawing any conclusion about the nature of either spin fluctuation spectrum. The two-fluid description makes this possible. Above $T^*$, $T_1^{-1}$ measures the strength of the magnetic fluctuations of the unhybridized local moments while below $T^*$, the local moment contribution is reduced by their hybridization and a heavy electron component emerges, producing the change in the temperature dependence of the total $T_1^{-1}$ that has been seen in many other heavy electron materials \cite{Curro2009}. Assuming no interference between the local and heavy electron contributions to $T_1^{-1}$, the two-fluid model suggests that it takes the following form:
\begin{equation}
\frac{1}{T_1}=\frac{1-f(T)}{T_{1l}} + \frac{f(T)}{T_{1h}},
\label{Eq:T1}
\end{equation}
where $f(T)=(1-T/T^*)^{3/2}$ is the order parameter and represents the fraction of the local moments that are converted into the itinerant Kondo liquid. The local term $T_{1l}$ is from the unhybridized local moments and its temperature dependence can be obtained by fitting to the high temperature $T_1$ above $T^*$ and making the assumption that this temperature dependence persists below $T^*$. The basis for our analysis is shown in Fig.~\ref{Fig:CoT1} where the experimental data take the form $T_{1l}^{-1}=a+bT$ above $T^*$, that has been seen for many heavy electron materials. It originates in the weak coupling between local moments that gives rise to their measured Curie-Weiss susceptibility, and its magnitude near $T^*$ scales with the strength of the nearest neighbor inter-ion exchange interaction, which is of order $T^*$ \cite{Yang2008b}. 

For CeCoIn$_5$ at different pressures, we estimate $T^*$ from transport experiments displaying a Hall anomaly \cite{Yang2008a}; we find $T^*\sim 50\,$K at ambient pressure, $\sim 60\,$K at $0.58\,$GPa and $\sim 70\,$K at $1.2\,$GPa.  We note that the NQR $T_1^{-1}$ contains the effect of both c and planar directions, so we should use an average  $T^*$ that is slightly larger than the planar one used for the planar Knight shift analysis due to anisotropy. Assuming no change in the hyperfine coupling with pressure, the inset of Fig.~\ref{Fig:CoT1} compares $T_1(T^*)$ and $T^*$ and confirms their proportionality.

\begin{figure}[t]
{\includegraphics[width=9cm,angle=0]{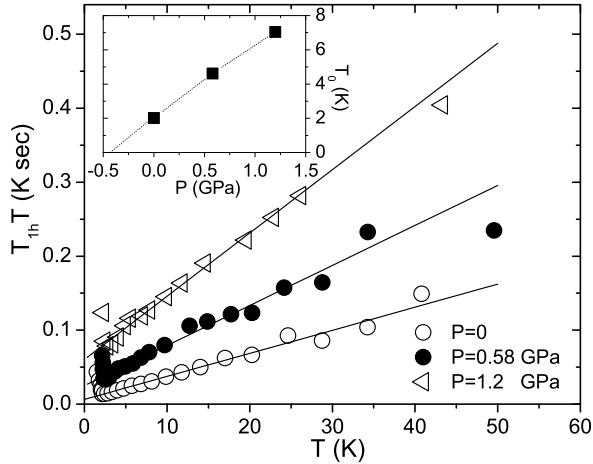}}
\caption{
{Kondo liquid $T_{1h}T$ for CeCoIn$_5$ at different pressures. The linear temperature dependence of $T_{1h}T$ signals the presence of quantum critical fluctuations. The inset plots the deduced $T_0$ that measures the distance from an antiferromagnetic quantum critical point.
}
\label{Fig:CoT1KL}}
\end{figure}

To obtain the heavy electron contribution to the relaxation rate, we assume that the intrinsic temperature dependence of the local moment contribution to $T_{1l}^{-1}$ extends to lower temperatures; what changes is its strength. The dotted lines in Fig.~\ref{Fig:CoT1} give the local moment contribution $[1-f(T)]/T_{1l}$. The Kondo liquid contribution can then be obtained by using Eq.~(\ref{Eq:T1}). The resulting $T_{1h}T$ are plotted in Fig.~\ref{Fig:CoT1KL}. Interestingly, this Kondo liquid relaxation rate fits well to the following simple formula for all pressures below 1.2 GPa,
\begin{equation}
T_{1h}T \propto (T+T_0)
\label{Eq:TH}
\end{equation}
where $T_0$ is a positive constant of the order of a few Kelvin. The slope increases with increasing pressure or $T^*$. If we take the above linear formula for the Kondo liquid as our starting point and fit to the experimental data, we do find a good fit (solid lines in Fig.~\ref{Fig:CoT1}). The scaling behavior in Eq.~(\ref{Eq:TH}) has been found in many high $T_c$ cuprates where it signals the presence of quantum critical fluctuations of a nearly 2d spin liquid \cite{Barzykin2009}. This provides additional evidence of the similarity between these two kinds of unconventional superconductors. The deviation from Eq.~(\ref{Eq:TH}) above 40 K arises either from experimental errors or the breakdown of the quantum critical behavior in that high temperature region. 

$T_0$ measures the distance from a quantum critical point. In the inset of Fig.~\ref{Fig:CoT1KL}, we plot $T_0$ as a function of pressure and conclude that the quantum critical point for this material is at a slightly negative pressure, consistent with previous expectations \cite{Sidorov2002}. The increasing strength of these quantum critical fluctuations $f(T)/T_{1h}$ of the Kondo liquid with lowering temperature for $T_0<T<T^*$ is in strong contrast to the overall decrease of $T_1^{-1}$, while these are a highly promising candidate for their superconductivity below $T_c$.

To summarize, we find scaling behavior for the Knight shift anomaly in both the normal and superconducting phases of the heavy electron superconductor CeCoIn$_5$ that provides a natural explanation for the various unusual features of the Knight shift data and reflects directly the heavy electron Kondo liquid condensation into the superconducting state. The quantum critical fluctuations we deduce from our two-fluid analysis of normal state behavior provide a highly promising candidate for the physical origin of their superconductivity, while the pressure variation of the calculated temperature dependence of the intrinsic heavy electron Kondo liquid spin-lattice relaxation rate suggests that the magnetic QCP is located in this material at a slightly negative pressure, as suggested by previous experiments. The theoretical method applied in this analysis can be easily extended to other heavy electron materials, and we find that on doing so for CeRhIn$_5$ that the magnetic quantum critical fluctuations seen there are likewise in the heavy electron channel, suggesting that its long range magnetic order is triggered by a spin density wave instability in this component.

Work at Los Alamos was performed under the auspices of the U.S. Department of Energy through the LANL/LDRD program. We thank our colleagues in the Aspen Center for Physics workshop on "Correlalted Behavior \& Quantum Criticality in Heavy Fermion and Related System" for stimulating discussions and the Aspen Center for Physics for its hospitality during the writing of the paper. Y.Y. thanks Joe Thompson for discussions and the ICAM Fellow program for its support. R.R. Urbano thanks the NSF Cooperative Agreement \# DMR-0654118 and the State of Florida.

\end{document}